\documentclass[acmsmall]{acmart}
\usepackage{algorithm}
\usepackage[noend]{algpseudocode}
\makeatletter
\usepackage{amsmath}
\usepackage{svg}
\def\algbackskip{\hskip-\ALG@thistlm}
\makeatother
\algnewcommand\algorithmicforeach{\textbf{for each:}}
\algnewcommand\ForEach{\item[ \algorithmicforeach]}
\algdef{S}[FOR]{ForEach}[1]{\algorithmicforeach\ #1\ \algorithmicdo}
\usepackage{caption}
\usepackage{subcaption}
\usepackage{array}
\usepackage{multirow}

\newcommand\MyBox[2]{
  \fbox{\lower0.75cm
    \vbox to 1.7cm{\vfil
      \hbox to 1.7cm{\hfil\parbox{1.4cm}{#1\\#2}\hfil}
      \vfil}%
  }%
}
\AtBeginDocument{%
  \providecommand\BibTeX{{%
    \normalfont B\kern-0.5em{\scshape i\kern-0.25em b}\kern-0.8em\TeX}}}

\setcopyright{acmcopyright}
\copyrightyear{2018}
\acmYear{2018}
\acmDOI{10.1145/1122445.1122456}

\acmJournal{JACM}
\acmVolume{37}
\acmNumber{4}
\acmArticle{1}
\acmMonth{8}

\begin{document}

\title{Spatiotemporal Data Mining: A Survey}


\author{Arun Sharma}
\email{sharm485@umn.edu}
\orcid{0002-6908-6960}
\affiliation{
  \institution{University of Minnesota, Twin Cities}
  \city{Minneapolis}
  \state{Minnesota}
  \country{USA}
}

\author{Zhe Jiang}
\email{zhe.jiang@ufl.edu}
\affiliation{%
  \institution{University of Florida}
  \city{Gainesville}
  \state{Florida}
  \country{USA}
}

\author{Shashi Shekhar}
\email{shekhar@umn.edu}
\affiliation{%
  \institution{University of Minnesota, Twin Cities}
  \city{Minneapolis}
  \state{Minnesota}
  \country{USA}
}

\renewcommand{\shortauthors}{Arun Sharma ,Zhe Jiang and Shashi Shekhar}

\begin{abstract}
Spatiotemporal data mining aims to discover interesting, useful but non-trivial patterns in big spatial and
spatiotemporal data. They are used in various application domains such as public safety, ecology,
epidemiology, earth science etc. This problem is challenging because of the high societal cost of spurious
patterns and exorbitant computational cost. Recent surveys of spatiotemporal data mining need update
due to rapid growth. In addition, they did not adequately survey parallel techniques for spatiotemporal
data mining. This paper provides a more up-to-date survey of spatiotemporal data mining methods.
Furthermore, it has a detailed survey of parallel formulations of spatiotemporal data mining.
\end{abstract}



\maketitle

\section{Introduction}
Spatiotemporal data mining is the process of discovering novel, non-trivial but potentially useful patterns in large scale spatiotemporal data. Spatiotemporal (ST) data include georeferenced climate variables, epidemic outbreaks, crime events, social media, traffic, transportation dynamics, etc. Analyzing and mining such data is of great importance for advancing the state-of-the-art in many scientific problems and real-world applications due to its interdisciplinary nature. Hence, ST data are used prominently in various domains such as public safety, ecology, epidemiology, etc. 




\begin{figure}[h]
  \centering
  \includegraphics[width=1.0\linewidth]{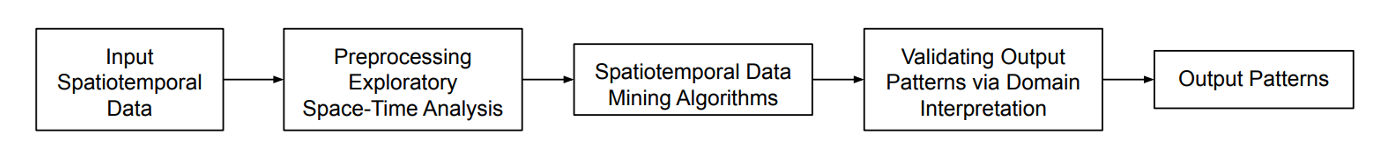}
  \caption{The Process of Spatiotemporal Data Mining (Best in color)}
  \label{fig:Flowchart}
  \Description{}
\end{figure}

Figure \ref{fig:Flowchart} shows the overall process of spatiotemporal data mining. After preprocessing to remove noise, errors, etc., the input data undergo space-time analysis to understand their spatiotemporal distribution. Appropriate spatiotemporal data mining algorithms are applied to produce output patterns that are then studied and validated by domain experts to discover novel insights, and the data mining algorithms are refined accordingly.  

\begin{figure}[h]
  \centering
  \includegraphics[width=1.0\linewidth]{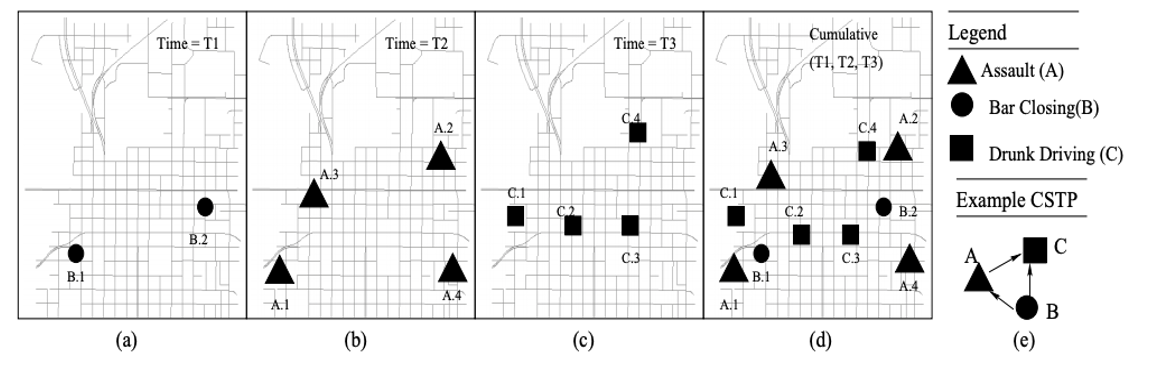}
  \caption{Illustrative ST Crime Dataset and Cascading ST Pattern. (Best in color)}
  \label{fig:STPattern}
  \Description{}
\end{figure}

An example of a spatiotemporal (ST) pattern is shown in Figure \ref{fig:STPattern}. The figure depicts a cascading spatiotemporal pattern \cite{mohan2010cascading} that was output on an urban crime dataset involving several event types (bar closings, drunk driving, assaults). Figures 2(a) -2 (c) show the instances of each event type with their times and locations, and Figure 2 (e) shows the same event instances altogether.  The detected ST pattern suggests the interesting hypothesis that bars at closing time are potential generators of drunk driving and assault crimes in nearby areas. Such information can help law enforcement agencies, public safety groups, and policymakers determine appropriate crime mitigation action.

Mining spatiotemporal data via algorithms pose unique computing challenges. For instance, due to large data volume, users may experience unacceptable runtime or expending great efforts in pre-processing the data due to great variety. Besides volume and variety, high data velocity relative to ingest capacity is another computing challenge where user experiences frequent data loss. For instance, the real-time processing of millions of tweets involves real-time pre-processing and mining of certain information. COVID 19 Safegraph data \cite{safegraph} records human mobility movements every minute via nearby information (e.g., POIs, Business Categories) across the country. This involves real-time updates on information such as the number of visits to different POIs resulting in massive data volume (Terabytes), which is impossible to store and process in a local machine. In addition, the US Census \cite{USCensus} including millions of records saved in different tables and schema spanning terabytes of data space.

Querying such data may be time-consuming, and further applying algorithms often perform poorly in terms of computation time at a large scale on scientific or geographic computing platforms. Such computations require a scalable and reliable software ecosystem to solve broad research problems and efficiently help society’s decision-makers.  Hence, it is important that a broad community of scientists and users be informed about the high performance, scalable, and open-source software tools available today that can facilitate spatiotemporal data analytics with parallel computation to significantly advance domain research.

However, most of these efforts have received little attention in recent surveys of spatial and spatiotemporal data mining research. Here we provide a comprehensive survey of spatiotemporal data mining techniques along with a brief description of their statistical foundations and major output pattern families (e.g., outliers, predictions, hotspots, etc.). We also provide current literature on recent approaches that are being widely studied in both sequential and parallel processing environments.

Similar to Shekhar et al. \cite{shekhar2015spatiotemporal}, this survey starts with a review of previous surveys, followed by a statement of our contribution towards this survey (Section 2). Section 3 provides key terms related to spatial and spatiotemporal data and describes their statistics foundation along with a brief statement about the societal importance of parallel processing in spatiotemporal data mining. Section 4 describes six main output pattern families (i.e., spatiotemporal outliers, telecouplings, prediction, partitioning and summarization, hotspots, and change detection) along with their respective detection approaches in both sequential and parallel frameworks. Some current research tools are discussed in Section 5, including state of the art parallel tools, frameworks, and libraries, which are being used in many applications. Section 6 concludes the chapter with a look at the current research trends and future directions.


\section{Related Work and Contribution}
\label{section:realted_work_contribution}
\subsection{Framework}
\label{section:FrameWork}
A large number of literature surveys can be found on spatial and spatiotemporal data mining. Older surveys focused on methodologies without statistical significance tests, whereas recent works have focused on pattern families with limited attention to parallel spatiotemporal data mining techniques. Hence, we categorized it into two groups, one without a statistical foundation and the other with statistical foundations, as shown in Figure 3.  The earliest of these date to the 1990s \cite{ester1997spatial,koperski1996spatial,roddick1999bibliography}. Miller et al. \cite{roddick1999bibliography} later reviewed trends in spatial and spatiotemporal data mining; like earlier surveys, none of the works covered considered statistical significance.  

\begin{figure}[ht]
    \centering
    \includegraphics[width=0.5\textwidth]{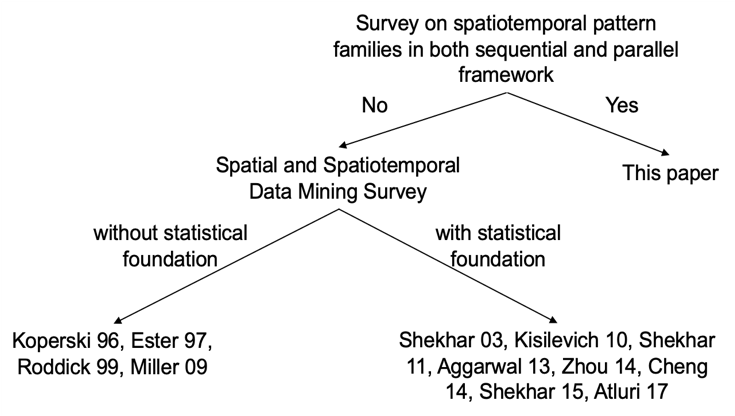}
    \caption{Categorizing Spatial and Spatiotemporal Data Mining Surveys\\}
    \label{fig:Survey}
\end{figure}

Shekhar et al. \cite{shekhar2003trends} focused on the unique characteristics of spatial data and offered the first in-depth review of its statistical foundation. Zhou et al. \cite{zhou2014spatiotemporal} described various techniques involved in change detection from an interdisciplinary perspective; Aggarwal et al. \cite{aggarwal2017introduction} reviewed spatial and spatiotemporal outlier techniques, and Cheng et al. \cite{cheng2014spatiotemporal} talked about spatiotemporal autocorrelation, space-time forecasting, space-time clustering and its visualization. More comprehensive reviews by Shekhar et al. \cite{shekhar2011identifying} covered most of the pattern families but did not include any research done in the parallel computing domain. Among recent works, Atluri et al. \cite{atluri2018spatio} give a good overview of spatiotemporal data mining that addresses pattern families and describes the methods used in a parallel framework to some extent. Overall, however, there is no current survey on spatiotemporal data mining that covers prominent pattern families from both a sequential and parallel perspective.

\textbf{Contributions:} This chapter makes the following contributions: (i)  It surveys major pattern families with current literature in spatiotemporal data mining from both sequential and parallel computing perspective; (ii) It informs researchers about the current tools, platforms, and libraries that are available for spatiotemporal data analysis in parallel computing environments and (iii) It summarizes recent trends and suggests future work for the advancement of high-performance spatiotemporal data mining.

\section{Definitions and Societal Importance}
\label{BasicConcept}
This section provides key terms and a taxonomy of spatial and spatiotemporal data and its attributes and relationships. It also explains why parallel computing in ST data mining is essential for society.

\textbf{Spatial and Spatiotemporal Data:} Spatial data provides information related to different instances in coordinate space and can be represented as object models (points, lines, and polygons), field models, and spatial networks (e.g., graphs) \cite{shekhar2007spatial}. Spatiotemporal data includes additional temporal information and can be represented by temporal snapshot models, temporal change models, and event or process models. In temporal snapshot models, snapshots can be considered trajectories of lines and polygons, including spatial layers (i.e., points or multi-points). Temporal change models represent spatiotemporal data with incremental changes from a given start time on a spatial layer (e.g., Brownian motion, random walk). Event models represent events as entities that remain consistent over time, while process models represent processes as entities that are subject to change over time.

\textbf{Spatiotemporal Data Attributes:} Attributes of spatiotemporal data include spatial and non-spatial characteristics and temporal attributes that include snapshots of spatial objects, raster layers, spatial networks, and process durations. Since materializing spatiotemporal relationships may result in the loss of spatial information, spatiotemporal statistics and techniques are preferred over traditional methods. Spatiotemporal statistics integrate spatial and temporal statistics based on first and second-order moments (mean, variance, co-variance) and can be further classified into descriptive and dynamic models \cite{cressie2015statistics}.

\textbf{Societal Importance:} Geo-spatial algorithms use spatial and spatiotemporal data in polygonal or raster form. However, these algorithms are hard to execute due to the irregular structure of the statistics' data and complexity, making it important to optimize computations for time-critical operations. In turn, runtime optimization supports timely decision-making by city planners, policymakers, federal governments, etc. For instance, real-time emergency response is required in a pandemic (COVID-19) or a natural disaster (e.g., hurricane, tornado) where every second lost in the decision-making process (e.g., rescue operation) may lead to property damage, injuries, or loss of life \cite{prasad2017parallel}.

\begin{figure}[ht]
    \centering
    \includegraphics[width=1.0\textwidth]{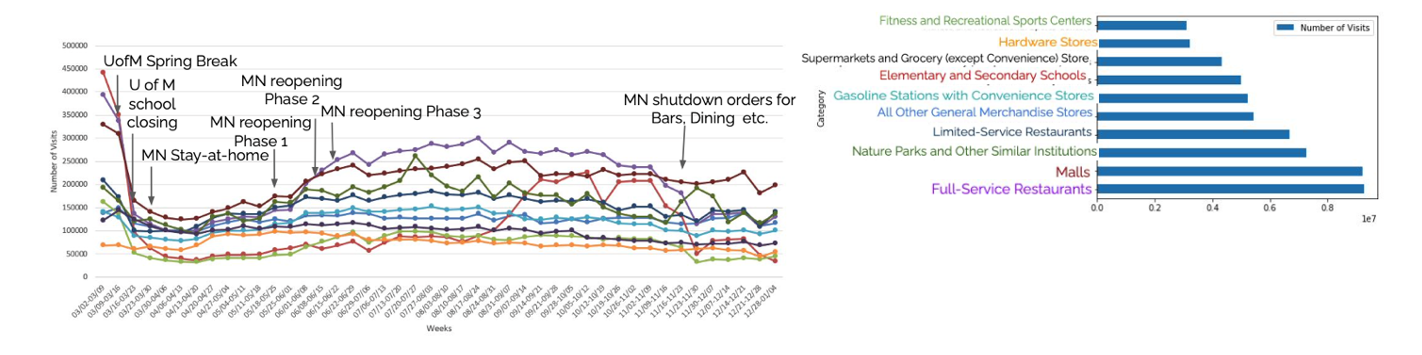}
    \caption{Most Frequently Visited Business Categories in Minnesota \cite{sharma2022understanding}\\}
    \label{fig:Survey}
\end{figure}

Our recent work \cite{Sharma2020}, we visually monitored human mobility patterns every week for every Point of Interest (POI) and Business Category in Minnesota during an early phase of the COVID-19 pandemic. Figure 4 shows the change in weekly mobility patterns for the most frequently visited business categories, positively correlated to several cases. This information helped Minnesota policymakers plan how to safely reopen the state economy while reducing the virus's spread. In another example, Chang et al. \cite{chang2021mobility} used US Census data to study the effects of mobility among disadvantaged racial and socioeconomic groups for more-effective and equitable policy responses to COVID-19.  However, computing such intricate mobility patterns involves large data volumes with millions of records updated periodically in real-time. Hence, big data processing platforms \cite{zaharia2012resilient} and hardware (e.g., graphical processing units) tools are used are needed to perform accelerated computing and provide output in real-time for time-critical decision making.

\section{Output Pattern Families}
\label{ProblemStatment}
This section provides a brief description of major pattern families with current literature surveys in non-parallel and parallel frameworks. This section describes six major pattern families in spatiotemporal data mining and briefly surveys the current related literature in non-parallel and parallel environments.

\subsection{Spatiotemporal Outliers and Anomalies}
\label{Baseline}
Outliers are sets of observations that appear to deviate from expected behavior as compared to other observations, resulting in abnormal patterns that may arouse suspicion in some cases (e.g., illegal fishing \cite{sharma2020analyzing}). A spatial outlier is a spatially referenced object whose non-spatial attribute value significantly varies from other spatially referenced objects in the spatial neighborhood.  Figure 5b shows a variogram cloud of the example dataset shown in Figure 5a, where the locations which are near one another but with large attribute differences might indicate a spatial outlier. For instance, pairs (P, S) and (Q, S) in Figure 5b may be potential spatial outlier candidates since they lie above the main group of pairs. Point S might also be considered a spatial outlier since it is present in both the (P, S) and (Q, S) groups. A spatiotemporal outlier that differs significantly both spatially and temporally from its spatiotemporal neighborhood results in instability or inconsistency of the actual data. Statistics include bi-partite multidimensional tests (e.g., Moran scatterplot) and quantitative tests (e.g., scatterplots), which can also be applied to spatiotemporal data.

\begin{figure}[htb]
    \centering 
\begin{subfigure}{0.35\textwidth}
  \includegraphics[width=\linewidth]{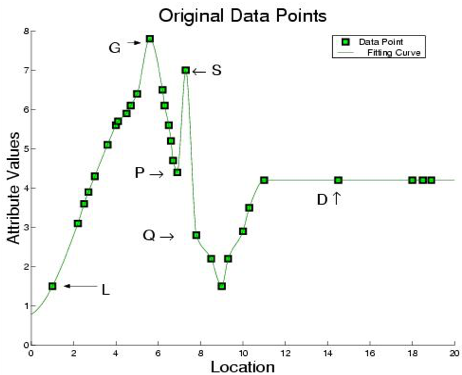}
  \centering
  \captionsetup{justification=centering}
  \caption{An Example Dataset}
  \label{fig:Input}
\end{subfigure}\hfil 
\begin{subfigure}{0.35\textwidth}
  \includegraphics[width=\linewidth]{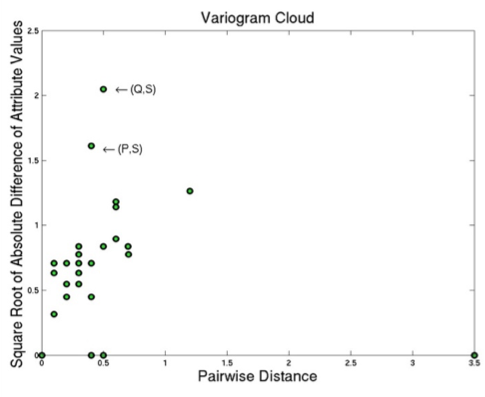}
  \centering
  \captionsetup{justification=centering}
  \caption{Variogram Cloud Scatterplot\\}
  \label{fig:Output}
\end{subfigure}\hfil 
\centering
\captionsetup{justification=centering}
\caption{Example Dataset and Variogram Cloud Scatterplot \cite{shekhar2003trends} (Best in color).}
\label{fig:Variogram}
\end{figure}


\subsubsection{Common Approaches}
\label{baseline1}
Outliers in spatial time series: Detection of spatial time series outliers is based on visualization approaches (e.g., variograms, Moran scatterplots) and the spatiotemporal neighborhood, which can be identified with distance measures such as K nearest neighbors or graph connectivity \cite{banerjee2003hierarchical}. The methods can also be extended to include non-spatial, weighted, categorical, or local spatial outliers, etc.
In the parallel computing domain, a tile-map interface is proposed for detecting anomalous events \cite{shin2017stexnmf} by simultaneously processing multiple tiles using multi-threading, providing substantial computational speedup. Another spatiotemporal outlier detection method \cite{barua2007parallel} uses a parallel wavelet transform to process massive meteorological data on high-performance architecture.\\

Anomalies in Spatiotemporal Data: Outliers in spatiotemporal data such as trajectories can be defined as anomalous patterns or anomalies. Many sequential approaches are used in finding outliers while monitoring trajectories \cite{zheng2015trajectory,lee2008trajectory}. Recently, Lam et al. \cite{lam2016concise} proposed a neighbor search approach to detect abnormal driving patterns that efficiently summarize and analyze their spatial and temporal patterns.  A trajectory partitioning strategy was recently proposed \cite{lu2017distributed} in the parallel domain using similarity measures and taking into account the multi-motion characteristics of a vehicle (e.g., speed, acceleration, etc.)\\

\subsection{Spatiotemporal Couplings and Tele-couplings}
\label{PotentialIntersection}
Spatiotemporal coupling patterns are based on instances that occur in close spatial and temporal proximity. These patterns can be unordered (co-occurrences), partially ordered  (cascading patterns), or ordered (sequential patterns). The finding that bar closings lead to drunk driving and assault, shown in Figure 2, is an example of a partially ordered cascading pattern.  In addition, there is the spatiotemporal tele-coupling pattern, which identifies a significant positive or negative correlation in a spatial time series.  Statistics include the spatiotemporal cross K function, an extension of Ripley’s K function with multiple variables.

Discovering various spatiotemporal coupling patterns and tele-coupling is important in applications related to ecology, environmental science, public safety, and climate science. Recent work proposed a data-driven approach for inferring traffic cascading patterns over a real-world dataset \cite{liang2017inferring}.

\subsubsection{Common Approaches}
\label{baseline3}
o-Occurrence Patterns: Spatiotemporal co-occurrence patterns are simultaneous occurrences of two or more subsets of events in both spatial and temporal proximity. Mixed drove co-occurrence pattern detection \cite{celik2006mixed} is one such approach used in various fields such as transportation planning, battlefield strategy, and gaming but these patterns are computationally expensive. Other work includes a monotonic composite interest measure \cite{pillai2013filter}. In parallel computing, asynchronous co-occurrence patterns \cite{yu2015spatio} have been studied for flood prediction by relating precipitation with actual precipitable water through association mining.  The framework uses a MapReduce paradigm for parallelization and scalability and identifies spatial associations on climate data between one target location and its corresponding asynchronous locations. 

Sequence Patterns: Sequence patterns are spatiotemporal events in the form of a “chain reaction” of subsequent ordered events. Some application domains include epidemiology to follow disease transmission patterns between several species. Methods proposed in \cite{huang2008framework} use a spatial sequence index as a significance measure that interprets the K-function. Other approaches \cite{zheng2011probabilistic} mine uncertain sequences via probabilistic methods.
Parallel implementation of sequential pattern mining \cite{qiao2008partspan} was proposed for trajectory data by first reducing candidate trajectories, performing data parallelism for counting support, and finally using a MapReduce model for distributing jobs for scalability. A similar MapReduce implementation \cite{liang2015sequence} proposed constructing a lexicographical sequence tree for efficiently finding frequent itemsets and then implementing a breadth-wide support pruning strategy for scalability over trajectories.   

Spatial time series and teleconnections: Teleconnections are used to identify pairs of spatial time-series at different locations and are important in climate science and other domains for understanding oscillations (e.g., pairwise correlations, spatial autocorrelation on spatial time-series). Finding these patterns is computationally expensive due to the length of the time-series and the need for frequent enumeration of many candidates. Other techniques \cite{zhang2003correlation,kawale2012testing} use spatial autocorrelation and statistical significance testing to reduce recurrent computations. DStep framework \cite{kendall2011simplified} was proposed for efficient domain traversal over data-intensive analysis tasks such as teleconnection analysis on atmospheric CO2 and climate data.

\subsection{Spatiotemporal Prediction}
\label{section:ProposedApproach}
Spatiotemporal predictions \cite{zheng2015trajectory} aim to learn a model that can forecast a target variable from the given explanatory feature's forecast-dependent variables. This includes predicting continuous values via spatiotemporal regression or classifying outcomes via spatiotemporal classification. Spatiotemporal prediction allows spatiotemporal dependency over variables (e.g., neighborhood relationship, strength between elements etc.) at different locations.

Applications vary from remote sensing, where features include spectral bands and dependent variables such as forest, urban, water, etc. Other applications include regression analysis to predict global or regional climate variables, real estate pricing, etc.  

\subsubsection{Common Approaches}
\label{TimePrioritizer}
Spatiotemporal Regression: More recently, deep learning has been shown to solve complex non-linear problems. Deep learning techniques have been applied in traffic flow prediction \cite{chen2018exploiting,moosavi2019accident} and duration of traffic incidents \cite{fu2019titan} by utilizing data attributes such as traffic events, weather data, points of interest, etc.
In the parallel computing domain, a novel spatiotemporal recurrent convolutional neural network (RCNN) is proposed \cite{yu2017spatiotemporal} in transportation science where the spatial dependencies are captured in deep CNN layers while the temporal dynamics are learned by Long Short Term Memory units and further applied in GPUs for accelerating the model learning procedure. A similar technique, DeepTC, uses a ConvLSTM network \cite{kim2018deeptc} for forecasting accurate tropical cyclone trajectories using a Weather Research and Forecasting Model. 

Spatiotemporal Kriging: Kriging is used for interpolating unknown observations based on prior knowledge of known trajectories. Semantic Kriging (SemK) \cite{bhattacharjee2016prediction} fetches land-use/land-cover distribution of terrain and incorporates the existing interpolation process of climatological pattern analysis. A parallel spatiotemporal Kriging algorithm \cite{zhang2018implementation} predicts points with increasing observations, varying acceleration proportional to prediction points, comparing the original point, and cross-validating with real-world temperature data with the traditional model.

\subsection{Spatiotemporal Partitioning and Summarization:}
Spatiotemporal partitioning is a method of dividing underlying space and time and clustering a set of similar observations by partitioning the. In contrast, a spatiotemporal summarization is a brief representation of the spatiotemporal data associated with a spatiotemporal partition where aggregated statistics represent each partition. An example of clustering a common sub-trajectory from a set of trajectories T1, T2, T3, T4 (as described in \cite{lee2007trajectory}) is shown in Figure 6. Scan statistics include spatiotemporal point density estimation and temporal correlation.

\begin{figure}[ht]
    \centering
    \includegraphics[width=0.45\textwidth]{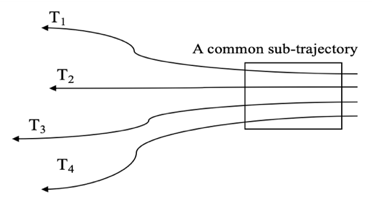}
    \caption{Trajectory Clustering\\}
    \label{fig:TrajectoryClustering}
\end{figure}

Spatiotemporal partitioning and summarization are important in application domains such as public health, public safety, environmental science, and law enforcement. For instance, partitioning vehicle accident data, which is spatial and temporal, helps law enforcement agencies find trends and effectively deploy their resources.

\subsubsection{Common Approaches}
Spatiotemporal Event Partitioning: Methods for partitioning spatiotemporal events can be classified as global density-based (e.g., DBSCAN, ST-DBSCAN); hierarchical (e.g., BIRCH \cite{zhang1996birch}), which partitions spatiotemporal data at different hierarchical levels; and graph-based (e.g., CHAMELEON \cite{karypis1999chameleon}), which proposes a sparse k nearest neighbor that partitions the graph into segments and later merges its fragments based on a similarity measure. A recent work \cite{xie2019significant} added a statistical significance test to DBSCAN that detects statistically significant clusters of various shapes and densities.

In parallel computing, balanced K-means \cite{von2018balanced} applies a new geometric based optimal partitioning algorithm, resulting in faster convergence. A parallel implementation of k-medoids clustering \cite{song2017pamae} uses a two-phase seeding and refinement approach that provides high accuracy and efficiency. DBSCAN and other density-based clustering methods \cite{welton2013mr,he2011mr,cordova2015dbscan} have been implemented across various frameworks and hardware. Spatiotemporal graph and hypergraph partitioning models have also been proposed on multicore architectures over sparse matrix data \cite{abubaker2018spatiotemporal}.

Spatial Time-Series Partitioning: Spatial time-series partitioning divides the space into identical regions where correlations between time-series within a region are maximum. A filter and refine approach \cite{zhang2003correlation} reduce computationally expensive and redundant computations.  In medical imaging, Fast-GPU-PCC has been proposed on functional MRI images \cite{eslami2018fast}.

Trajectory Data Partitioning: A recent survey \cite{huang2008framework} categorized trajectory partitioning methods based on time interval, the trajectory's shape, and semantic meaning. Another recent work \cite{jiang2017feature} partitions trajectories based on features such as density, speed, and direction flow. In the parallel domain, recent work \cite{bao2016managing} based on temporal partitioning using distributed and parallel RP-DBSCAN \cite{song2018rp} on Apache Spark uses a random and simultaneous partition scheme and find density reachable relationships and finally merges into a global cluster.

\subsection{Spatiotemporal Hotspots}

Given a set of geolocated objects, spatial hotspots are areas or regions where concentration of objects inside the region is significantly greater than outside. The log-likelihood ratio (Log LR) provides an interest measure for estimating the concentration of activities inside the given arbitrary shape (e.g., circle, ring) concerning the number of activities outside the shape in a given study area. Hence, the ratio is directly proportional to the concentration inside the given shape or region. It is complimented by a statistical significance test (e.g., p-value) that removes patterns that are highly unlikely to occur and is inversely proportional to the concentration of activities inside the region.

\begin{figure}[htb]
    \centering 
\begin{subfigure}{0.30\textwidth}
  \includegraphics[width=\linewidth]{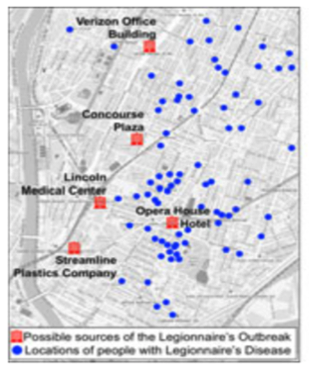}
  \centering
  \captionsetup{justification=centering}
  \caption{Legionnaire’s in New York}
  \label{fig:New York}
\end{subfigure}\hfil 
\begin{subfigure}{0.30\textwidth}
  \includegraphics[width=\linewidth]{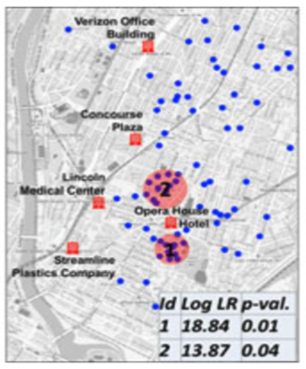}
  \centering
  \captionsetup{justification=centering}
  \caption{Circular Hotspots Output\\}
  \label{fig:Output}
\end{subfigure}\hfil 
\begin{subfigure}{0.30\textwidth}
  \includegraphics[width=\linewidth]{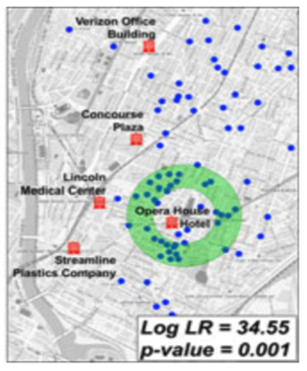}
  \centering
  \captionsetup{justification=centering}
  \caption{Ring Shape Hotspot Output\\}
  \label{fig:Output}
\end{subfigure}\hfil 
\centering
\captionsetup{justification=centering}
\caption{Circular and Ring-Shape Hotspot on New York Bronx Legionnaire Outbreak (Best in color)}
\label{fig:Hotspot}
\end{figure}

Figure 7 shows a case study of a New York Bronx Legionnaire’s disease outbreak \cite{eftelioglu2014ring} where the locations of 78 infected individuals (circular blue) and five potential outbreak sources (rectangular red) were hand digitized on a map (Figure 7a). Figure 7b shows a circular hotspot (in red) output by SatScan around the Opera House Hotel with a high concentration of geo-located infected individuals inside the circle along with their associated Log LR values and p-values. Figure 7c shows a ring-shaped hotspot output by a different method around the same location with a higher concentration (Log LR 34.55) of individuals inside the ring area (green) and better statistical significance (i.e., p-value 0.001).

Spatiotemporal hotspots are high density-based cluster patterns where the number of objects is unexpectedly more elevated than other observations within a specific time interval. Spatial hotspots (e.g., circular \cite{eftelioglu2014ring}, ellipsoidal \cite{kulldorff1997spatial}, linear \cite{tang2017significant}) can be extended to spatiotemporal hotspots by adding the dimension of time. Such hotspots are useful in application domains such as public health and criminology.

\subsubsection{Common Approaches}
Spatiotemporal Scan Statistics: Spatiotemporal hotspots can persist or emerge (e.g., disease outbreak) over time. Recently, Spatio-Temporal Network Kernel Density Estimation (STNKDE) \cite{romano2017visualizing} provides visualization for capturing the temporal dynamics of hotspots on network space and can be used for detecting traffic accidents. Other visualization techniques \cite{lukasczyk2015understanding} are based on the topological notion of temporal hotspots evolution over epidemiological and crime data. Lagrangian representation of linear hotspots is discussed over traffic data \cite{li2020significant}.
Parallel spatiotemporal hotspot detection was recently implemented using Getis-Ord G* as the scan statistic on a Spark framework \cite{mehta2016spatio} using a MapReduce paradigm. A novel normalization-based scan statistic was recently proposed \cite{xie2019nondeterministic} for robust spatial hotspot detection, which can also be extended to spatiotemporal hotspots by adding the time dimension.

\subsection{Spatiotemporal Change}
Spatiotemporal change accounts the change in footprint patterns depending on input data representations. A change is defined as a change in statistical distribution where the data is assumed to be a part of a certain distribution. A spatiotemporal footprint comprises both spatial and temporal direction where a temporal footprint can be classified as a single or set of snapshots and a point or an interval in a long series. A set of snapshots can represent a change in a spatial field whereas a single snapshot represents a purely spatial change. 

Other spatial footprints are raster based (local, focal, zonal) and vector based (point(s), polygon(s), line(s), network). A focal change occurs between a location and its spatial neighborhood while a zonal change occurs inside a spatial zone where a transition of data attributes occurs [10].  Figure 8. shows an example of spatial zonal change footprint of vegetation cover in Africa and its spatial zonal change patterns with longitudinal changes throughout the years.

\begin{figure}[htb]
    \centering 
\begin{subfigure}{0.50\textwidth}
  \includegraphics[width=\linewidth]{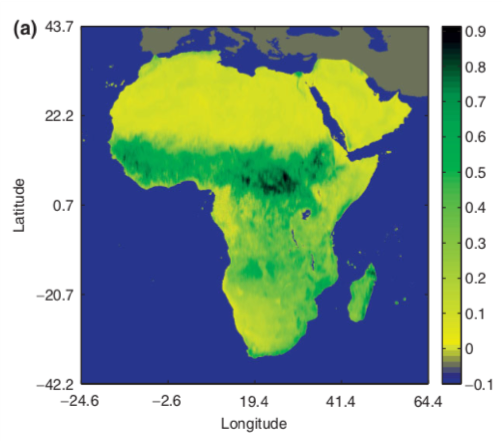}
  \centering
  \captionsetup{justification=centering}
  \caption{Vegetation Cover, August, 1981}
  \label{fig:Vegetation Cover}
\end{subfigure}\hfil 
\begin{subfigure}{0.43\textwidth}
  \includegraphics[width=\linewidth]{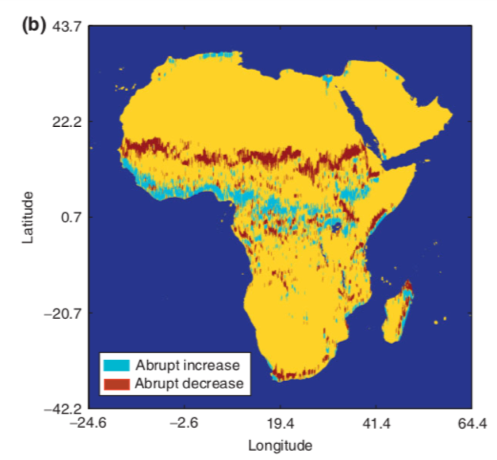}
  \centering
  \captionsetup{justification=centering}
  \caption{Spatial Zonal Change Patterns in Vegetation Cover of Africa with Longitudinal Changes\\}
  \label{fig:Output}
\end{subfigure}\hfil 
\centering
\captionsetup{justification=centering}
\caption{Spatial Zonal Change Footprint in Africa (best in color)}
\label{fig:ZonalChange}
\end{figure}

\subsubsection{Common Approaches}
Spatiotemporal Change Patterns with Raster-Based Spatial Footprints: A change in footprint can be a change in human activities, a natural disaster, climate change, etc., that is seen while comparing a collection of pixels between two raster snapshots. A change can also be zonal if we consider objects rather than pixels between snapshots \cite{zheng2015trajectory}. Scalable probabilistic change detection was proposed \cite{hong2016scalable} for satellite imagery using a parallel multicore architecture by parallelizing Gaussian and KL-Divergence computations.  Another parallel implementation uses recurrent convolutional neural network (RNN) architecture \cite{mou2018learning} for extracting both spectral and spatiotemporal features.

\section{Tools for Parallel Spatiotemporal Analysis}
The following provides an overview of the tools and libraries that are available for parallel spatiotemporal analysis, using both CPU and GPU resources as well as big data platforms such as Apache Spark. 

GIS Software: ArcGIS and QGIS are widely used GIS software packages that provide in-depth spatial and spatiotemporal analysis with the help of Parallel Processing Factor, which divides the input for performing operations on multiple processors. ArcGIS provides parallel extensions on various tools such as 3D Analyst tools, spatial analysis, space-time pattern mining etc. The popular QGIS analysis tool also provides OpenCL and CUDA support for parallel processing.

Spatiotemporal Statistics Tools: Many statistical packages (e.g., gstat, geoR) for spatial and spatiotemporal analysis can be run in the free R software environment but are limited to parallel implementation. The pbdNCDF4 \cite{patel2013quick} package provides parallel read and write capabilities for NetCDF datasets. Python’s deep learning Tensorflow library provides various functions for analyzing spatial and 3D spatiotemporal data.

Spatiotemporal Big Data Platforms: Big data systems are widely used for handling large complex data and exploiting parallelism on modern hardware. SpatialHadoop \cite{eldawy2015spatialhadoop} and GeoSpark \cite{yu2015geospark} provide support for spatial big data analytics. STARK \cite{hagedorn2017stark}, Geomesa \cite{hughes2015geomesa}, and webGlobe \cite{sharma2018webgiobe}, all based on the Apache Spark framework, integrate support for both spatial and temporal data operations.

\section{Research Trends and Future Needs}

This chapter covered state-of-the-art approaches for detecting six important spatiotemporal pattern families in both sequential and parallel computing environments. Our survey reveals that detection methods for some complex pattern families such as telecouplings (e.g., cascading patterns), summarization, change detection (e.g., vector-based footprints), etc., have not received much attention in parallel computing environments and can be a potential direction for future research.  

Big data tools that can leverage the resources of modern architectures for parallel processing also offer promising research opportunities. Current big data platforms such as HadoopGIS, Spatial Hadoop etc. provide distributed frameworks for parallel processing and some recent efforts towards processing spatiotemporal data such as trajectory data from open-source communities \cite{shang2018dita,ding2018ultraman} have a great potential to be extended toward finding non-trivial pattern families described in this chapter.

Besides parallel computing, other research directions might be interesting in spatiotemporal data mining. For instance, a space-time prism has been proposed \cite{miller1991modelling} for finding regions of uncertainty within trajectory gaps given the maximum speed of the object. Further exploration of the space-time prism \cite{kuijpers2017kinetic} considers multi-attributes such as acceleration and cases such as a sudden change in direction, providing more realistic path scenarios\cite{sharma2020analyzing,sharma2022analyzing}.

\begin{acks}
We would like to thank Kim Koffolt and the members of the University of Minnesota Spatial Computing Research Group for their comments.
\end{acks}

\bibliographystyle{ACM-Reference-Format}
\bibliography{manuscript}

\end{document}